# Dynamic Control of Local Field Emission Current from Carbon Nanowalls

Ying Wang, Yumeng Yang and Yihong Wu[a]

Information Storage Materials Laboratory, Department of Electrical and Computer Engineering, National University of Singapore, 4 Engineering Drive 3, Singapore 117583

[a] Electronic mail: elewuyh@nus.edu.sg

We report on a systematic study of modulation of the field emission current from carbon nanowalls using a sharp probe as the anode in an ultrahigh vacuum system. Modulation of the local emission current was achieved by either varying the anode-cathode distance ($d$) with the aid of an AC magnetic field or superimposing a small AC bias on a DC bias during the field emission measurement. Current modulation ratio of over two orders of magnitude was achieved with the modulation becoming more efficient at a smaller $d$. The experimental results are discussed using the Fowler-Nordheim theory in combination with a simple cantilever model to account for the modulation effect. The experimental results demonstrated good static stability and dynamic controllability of local field emission current from the carbon nanowalls.

# I. INTRODUCTION

Vertically aligned two-dimensional (2D) carbon with self-supported network structures, such as carbon nanowalls (CNW) or carbon nanosheets (CNS) have drawn much attention as potential emitter materials for nanoscale field emission devices due to their large height-to-thickness ratio, rigidity and endurance.[1-3] So far, various experimental efforts have been made to improve the field emission characteristics (such as turn-on electric field and stability of emission current *etc*.) of CNW/CNS; these include but are not limited to (i) reducing the screening effects among adjacent CNW/CNS flakes through selective growth,[4-8] (ii) improving the structure and morphology of CNW/CNS *via* fine tuning of the synthesis conditions, such as the types of carbon feedstock,[9] gas flow ratio,[10-13] deposition temperature,[14] substrate temperature,[12] and growth time,[13] (iii) chemical doping to reduce the turn-on field,[14-17] and (iv) surface treatment to improve the field emission characteristics of the as-grown CNW/CNS, such as selective coating of a thin layer of $Mo_2C$,[18] Au, Al and Ti,[19] plasma surface modification[20] and thermal desorption of absorbed hydrocarbons.[21] Most of the experimental results can be successfully explained by the Fowler-Nordheim (F-N) model[22] which predicts a linear relation between emission current (I) and applied electric field (E) in the F – N plot [i.e. $\ln(I/E^2)$ vs. $1/E$], though slight modification is sometimes needed to better account for the experimental observations. So far, very low turn-on field (i.e. the macroscopic electric field for an emission current density of 10 $\mu A/cm^2$) in the range ~0.23 – 6 V/µm has been reported on large-area samples (typical sample area larger than 1 $mm^2$) using a parallel plate configuration.[9,10,13,23-28] A stable milliampere-level field emission current for a duration of 1 – 200 hours has been achieved with both *d* and macroscopic applied electric field being kept constant. These results demonstrate the great potential of CNW/CNS as an efficient electron emitter for various applications. In addition to field electron emission sources, nanosized carbon emitters may also find applications in nanoscale vacuum electronic devices. For the latter purpose, in addition to static stability, good controllability over the emission current in a large dynamic range is also of crucial importance, such as those demonstrated in the gated field emitter design.[27,29,30] Considering the fact that practical nanoscale vacuum electronic devices are to be based on electron emission from nanosized emitters with the anode-cathode distance in the nanometer range, it is of great importance to study both the static and dynamic emission characteristics of CNW/CNS in an experimental configuration which resembles the actual device design and at the same time allows to perform the experiments in a controllable fashion. In this sense, the nanoprobe setup reported in our previous work is an ideal platform to carry out the intended studies.[31,32]

In the previous work, we have investigated systematically the relation between turn-on field and the anode-cathode distance for localized field emission from CNW/CNS samples. In this work, we study the dynamic properties of local field emission current from the CNW *via.* three different approaches. In the first approach (or Approach I), we used an *in-situ* AC magnetic field to periodically alter the distance between the CNW cathode and a sharp magnetic anode (i.e. a Ni probe) under a constant bias voltage. A schematic illustration of the experimental setup and energy diagram are shown in Fig. 1(a) and (d), respectively. In the second approach (or Approach II), a small AC modulating bias is superimposed on a DC bias to modulate the overall voltage bias across the anode-cathode gap [Fig. 1(b) and (e)]. This provides a variable macroscopic electric field between 0.3 kV/μm and 3 kV/μm in the direction along the emission gap. As a variation of Approach I (hereafter referred to as Approach III), the magnetic tip is replaced by a nonmagnetic one, and instead, the CNW is *in-situ* coated with a thin layer of Fe [Fig. 1(c) and (f)]. As we will discuss Sec. III, in Approach I and III, current modulation is mainly achieved through varying the anode-cathode gap. As for Approach II, although both bias voltage and anode-distance variations are expected to play a role, experimental results suggest that distance variation induced effect is dominant. The first two approaches are effective in modulating the emission current by over two orders of magnitude and, more importantly, the emission current is stable during the entire duration monitored.

## II. EXPERIMENTAL

### *A. CNW and probe preparations*

The CNWs were grown on Cu substrates using microwave plasma-enhanced chemical vapour deposition (MPECVD). The gases used were a mixture of $CH_4$ and $H_2$ with typical flow rates of 40 and 10 sccm, respectively. Before the $CH_4$ gas was introduced to the quartz tube to commence the growth of nanowalls, the substrate was pre-heated to about 650 – 700 $^o$C (limited by the microwave power) in hydrogen plasma with a bias of 50 V for 10 – 15 min. The typical growth time was 1 to 5 min. Details about the growth conditions, morphology and structural properties of the as-grown CNWs can be found elsewhere.[1,24] For local field emission measurements, sharp W and Ni probes were used as the anode. These probes were fabricated by electrochemical etching of W and Ni wires in NaOH and KCl solution (2M), respectively, using the drop-off lamellae method.[33] During each round of etching process, two probes are formed above and below the electrolyte lamellae, respectively. The lower probe with a larger taper length is always used for the present work. The Ni or W probes were loaded into the vacuum chamber immediately after the preparation to minimize air exposure. The CNW/Cu sample was fastened onto a sample holder which itself forms part of an *in-situ* electromagnet that is discribed below.

## B. Experimental setup for field emission

All the field emission measurements were performed in an Omicron UHV nanoprobe system with a base pressure better than $2.2\times10^{-10}$ mbar at room temperature. The nanoprobe system is equipped with four independently controllable nanoprobes and each probe module uses a piezo-electric inertia drive to achieve step motion with nanometer precision. Furthermore, the auto-approach capability of the probes ensures safe and non-destructive approach of probe to the sample surface. The whole measurement system is installed on a vibration isolation table using air legs, which itself is placed on the ground floor of a building to further minimize external disturbances. All these are critical for achieving precise control of *d* in the field emission measurements. The *in-situ* scanning electron microscope (SEM) allows for site specific field emission measurements down to nanometer scale. The sample stage is fitted with an electromagnet which is able to supply a vertical field up to 2000 Oe near the sample surface. The field can be controlled by an external bipolar power supply (Keithley 6221). Fig.2 shows a photo of the sample stage and schematic of the probe and electromagnet setup. All measurements were carried out in a Labview-based program which synchronises all source meters and allows real-time monitoring of the field emission current.

## C. Calibration of probe step height

Prior to field emission measurements, calibration measurements were performed to determine the step size of the probe (i.e. anode) using gold pads of different heights (0.2 – 2 μm), formed on a flat and heavily doped silicon substrate using standard optical lithography (Fig. 3). The exact height (h) of these patterned structures was measured by atomic force microscope. The detailed calibration procedure is as follows. Firstly, a sharp probe was first approached (or lowered) to tunneling regime of the surface of a gold pad using the auto-approach function of the nanoprobe controller (Step 1 in Fig. 3). After a high-resistance electrical contact was achieved, the feedback loop of the controller was de-activated to allow manual control of the probe position. The probe was then lowered further manually while the differential contact resistance is closely monitored until an ohmic contact was formed between the probe and the gold pad. Secondly, the probe was manually moved horizontally (< 2 μm) to above the trench between the gold pads, upon which the probe was lowered step-by-step at a preset speed till a contact with similar differential resistance was achieved (Step 2 in Fig. 3). The total number of steps (N) was recorded. The step size for downward probe motion was then calculated as $S_{down} = h/N$. The above steps were repeated for many times (> 15) to obtain the average downward step size ($<S_{down}>$).

The lifting or upward step size ($S_{up}$) was obtained by lifting the probe step-by-step at a preset speed for a certain step number ($N_{up}$) and then bringing it back into contact with the silicon substrate with a total number of steps ($N_{down}$); this gave the step size for upward motion: $S_{up} = N_{down} \times <S_{down}>/N_{up}$. This process was then repeated for many times to obtain the average lifting step size ($<S_{up}>$), which turned out to be ~1.38 nm/step for the chosen speed.

### *D. Shape formation of anode probe*

A ball-shaped probe (W or Ni) of desired size (100 nm – 2 μm) is subsequently prepared through a 3-step *in-situ* local electrical melting process by first applying a bias of appropriate amplitude between a sharp probe apex and the body of a relatively blunt W probe and then bringing them into contact for a self-limited discharge [Step 1 – 2 of Fig. 4(a)]. Upon formation of contact between the two probes, a closed current loop is immediately established and within a very short interval, the local heat generated by electrical discharge melts the apex of the anode probe into a sphere and automatically opens the circuit [Step 3 of Fig. 4(a)]. Figure 4(b) shows some typical SEM images of probes as prepared (sub-100 nm in apex size) and after (600 – 2200 nm) the local electrical melting process. It should be emphasized that this process is necessary to create a smooth anode surface without sharp protrusions, which is in turn crucial in determining accurately the anode-cathode distance and obtaining good reproducibility.

### *E. Procedures of performing local field emission measurements*

The blunt W probe was firmly pressed onto the CNW sample to form a low resistance electrical contact with the CNW cathode, while the anode was carefully approached to the top edge of a single CNW flake through monitoring the differential contact resistance using a lock-in amplifier setup.[31] After the an electrical contact between the CNW and anode was achieved, the anode was then lifted by a certain $N_{up}$ with the pre-calibrated $<S_{up}>$ to serve as an anode for field emission measurements at determined $d = N_{up} \times <S_{up}>$. Despite the fact that the edge of 2D carbon is not flat microscopically, emission of electrons occurs at 2D carbon sites that protrude along the direction of the applied electric field. These sites are expected to form contacts with the metallic anode first during the distance determination process described above. Thus, $d$ is naturally the distance between the emission sites and the anode bottom surface. A typical SEM image taken during a field emission measurement is shown in Fig. 4(c).

The measurements always began with ramping up the bias voltage till an emission current setpoint (typically 1 - 10 nA) was reached. The bias voltage was then kept constant for monitoring the emission current at a sampling rate of 8.3 Hz. The static stability of emission current was monitored in the first few hundred seconds. Upon reaching a stable emission, an external sinusoidal magnetic field $H = H_0 \sin t$ [for Approach I, Fig. 1(a)] or a small superimposed sinusoidal voltage bias $\Delta V = \Delta V_0 \sin t$ [for Approach II, Fig. 1(b)] was manually applied to examine the response of field emission current to the AC magnetic or electric field. Typical period of AC field is 15.3 s. The magnetization of all Ni probes was saturated in a large magnetic field along the emission gap direction prior to measurements.

After all measurements on bare CNW, the samples were *in-situ* evaporated with a thin Fe layer of a nominal thickness of 5 nm and 26 nm for Approach III. A large magnetic field along the emission gap direction was first applied to saturate the magnetization of the Fe coating. The same field emission measurements as in Approach I (except for the Ni probe replaced by a W probe) were then repeated on the Fe-coated CNW samples [Fig. 1(c)].

## III. RESULTS AND DISCUSSION

### *A. Stability of local field emission*

The local field emission measurements were monitored closely using the *in-situ* scanning electron microscope. No visible shift of the relative positions of the studied CNW flake and probe was observed. The good stability has further been confirmed by the current-time plot [Fig. 5(a)], in which the emission current stays very stable under static conditions throughout the whole monitoring process except for the initial training period of the first few tens of seconds. The spike at 454 – 456 s is presumably caused by isolated external disturbance after which the emission current recovered to its original value. The measurement was manually stopped after ~10 minutes, which is the typical duration used for one round of modulation measurements. Nevertheless, this does not mean that the emission current is only stable for this period of time. It should be noted that Fig. 5(a) is obtained at a much higher emission current (i.e. ~150 nA) than the typical current (< 10 nA) in the dynamic response study at zero modulation field. The purpose is to show that the emission current remains sufficiently stable even for a larger current than the pre-set sourcemeter compliance (i.e. 100 nA) during the current modulation investigation. For example, Fig. 5(b) shows a small stable local field emission current over a time span of over 40 minutes, obtained with the W probe shown in the inset.

It should be emphasized that screening effect from neighboring CNW flakes should be negligible since the spacing between adjacent flakes is normally ~1 μm, which is much larger than the investigated range of distance (i.e. ~1 nm < $d$ < 12.4 nm). If this was not the case, the measured field emission characteristics should depend on the anode size. To confirm this, we have performed field emission measurements on the CNW sample with different anode sizes ranging from 600 to 2200 nm. Considering that a $d$ larger than 150 nm requires a field emission ignition voltage higher than the maximum output voltage of the source meter (Keithley 2400), the emission current is predominantly coming from the CNW flake directly under the anode. As seen in Fig. 5(c), the relation between $d$ and the electric field required for an emission current of 1 nA obtained with different anode sizes closely overlap with each other. The increase of the required field with decreasing distance is due to a smaller field enhancement factor (defined as the ratio between the actual local electric field at the emitter surface and the macroscopic field) at a smaller $d$.[32] These results strongly show that the effect of neighboring flakes on both field distribution and emission current is negligible in our experimental setup.

## *B. Dynamic control of local field emission current with a Ni anode in an AC magnetic field*

Fig. 5(d) shows the typical response of the emission current to 15 cycles of an external sinusoidal magnetic field of amplitude $H_0$ = 80 - 158 Oe, obtained from the location shown in Fig. 4(c). The measurement was performed at zero-field distance $d_0$ = 11 nm (i.e., distance in zero magnetic field), and the relation between emission current and electric field at a fixed distance was found to be in good agreement with the F-N model, in consistence with our previous work.[32] Fig. 6(a) is a color contour plot of emission current as a function of time where the color scale has been normalized with respect to the zero-field current ($I_0$, defined as the emission current at t = 0 without magnetic field). Superimposed with the color contour plot are the emission current in one cycle of a sinusoidal magnetic field of amplitude $H_0$ = 37 Oe (white solid curve) and 136 Oe (black solid curve), respectively; the dotted line indicates the time when the emission current returns to $I_0$. When $H_0$ < 50 Oe, the time-dependence of emission current exhibits approximately a sinusoidal shape in the positive half-cycle but a rather flattened shape in the negative half cycle. This is a direct consequence of the combined effects of the variation of *d* caused by the magnetostatic interactions between the Ni probe and the applied magnetic field, and the exponential *d*-dependence of field emission current. In a weak applied magnetic field, the magnetization of the Ni probe is oriented along the probe axis direction due to strong shape anisotropy. Further, the magnetic flux in the probe is expected to be concentrated to the probe apex since it is magnetostatically unflavored for the magnetic flux to leak out from the side walls.[34] When a weak magnetic field is applied, the Ni probe is magnetostatically deflected downwards (upwards) depending on the direction of applied field, as illustrated in the upper (lower) inset of Fig. 6(b). In turn, the attracted (repelled) state of the Ni probe reduces (increases) the anode-cathode distance [Fig. 1(e)], resulting in a larger (smaller) emission current than $I_0$. To further elaborate this point, typical I-t curves corresponding to $H_0$ = 12.5 – 74.5 Oe were fitted using the F-N relation[22] with the local electric field replaced by $F = \beta V / [d_0(1 - \alpha \partial H_z / \partial z)]$:

$$I = \frac{Sa\beta^2}{\Phi}[\frac{V}{d_0(1-\alpha\frac{\partial H_z}{\partial z})}]^2 \exp(-\frac{b\Phi^{3/2}}{\beta}\frac{d_0(1-\alpha\frac{\partial H_z}{\partial z})}{V}), \quad (1)$$

where α is a constant in unit of nm Oe$^{-1}$ characterizing the strength of the probe-field interaction, S is the emission area (in the order of hundreds of nm$^2$), and a = $1.54 \times 10^{-6}$ AV$^{-2}$eV, b = 6.83 eV$^{-3/2}$Vnm$^{-1}$, Φ = 5 eV, V = 53.6 V, $d_0$ = 11 nm. β is the field enhancement factor which also depends on $d$.[32] However, the change of β is typically less than 1.5% for all investigated $H_0$ and contributes insignificantly to the observed change in the emission current. Thus, a constant β of 1.15 (calculated from the slope of the F – N curve at $d_0$ = 11 nm) has been used to simplify the following discussion. In Eq. (1), $\partial H_z / \partial z$ is the field gradient near the apex of the probe. In the specific magnet design used in this work, the gradient is approximately given by $5 \times 10^{-7}$ H (in unit of Oe nm$^{-1}$), where H (in unit of Oe) is the field strength at the top surface of the central magnetic pole. The experimental data (symbols) are plotted together with the optimum fitting curves (blue solid curves) in Fig. 6(c). All curves but the lowest one have been shifted vertically for clarity, and the figure beside each curve is the corresponding $H_0$ value in unit of Oe. It can be seen that the fitting results are satisfactory for $H_0 \leq$ 49.8 Oe. Furthermore, inset of Fig. 6(c) compares the extracted maximum deflections of the probe ($\Delta d = d_0\alpha(\partial H_z / \partial z)$) at different $H_0$ (symbols) and the simulation result (solid line) from a simple relation derived from a cantilever model:[35]

$$\Delta d = -(\sin\theta L)^3 \frac{\mu_0 M_s v \cos\theta}{3E_m I^*} \frac{\partial H_z}{\partial z}, \quad (2)$$

where $\mu_0$ is the permeability of free space, L = 8.5 mm is the probe length, θ = 45° is the angle between the probe axis and the normal of the sample surface, $I^* = 1.92 \times 10^{-4}$ mm$^4$ is the inertia, v ≈ $3.27 \times 10^{-2}$ mm$^3$ is the volume estimated from the probe shape and dimensions, $M_s = 5.12 \times 10^5$ A/m is the saturation magnetization of fcc Ni, and $E_m = 2.07 \times 10^{11}$ N/m$^2$ is the modulus of elasticity for Ni. Good agreement is obtained between experimental data and simulation results. It is worth mentioning that although the effect of magnetostriction cannot be ruled out completely, it does not play a significant role in modulating emission current in the present work mainly due to two reasons: (1) the negative magnetostrictive strain of polycrystalline Ni[36] would result in an increase of *d* and in turn a smaller emission current in an applied magnetic field, in contradiction to Fig. 6(b), and (2) the I-H dependence would be symmetric with respect to zero H if magnetostriction was the mechanism of the observed current modulation. The latter argument also excludes electron focusing as the dominant mechanism.

For $H_0$ > 62.1 Oe, a second peak in the emission current is observed (t = ~12 s) in the I-t curves [Fig. 6(c) and 6(d)]. The origin of this second peak can be understood more intuitively through the I-H curve shown in Fig. 6(e). When the amplitude of the magnetic field is sufficiently large, switching of the magnetization occurs in the Ni probe, leading to downward probe deflection at both positive and negative half-cycles of the AC magnetic field [upper insets in Fig. 6(e)]. Interestingly, a few fine features in the I-H curves are constantly observed due to the very sensitive exponential dependence of the current on *d*. Firstly, the steep increase in emission current in the range from ~-70 Oe to -100 Oe is corresponding to the reversal of the net magnetization in the Ni probe. One or a few small jumps before this magnetization reversal are constantly observed at ~-60 Oe. This can be seen more clearly in Fig. 6(f) where typical normalized I-H curves corresponding to different $H_0$ in the range from 87 Oe to 148 Oe are shown. These jumps suggest that the magnetization reversal of the Ni probe used in this work consists of reversal of some small domains followed by a rapid reversal of the magnetization of the entire probe. Secondly, the kink at ~-90 Oe is believed to indicate the completion of the reversal process of the net magnetization [Fig. 6(e)]. Further increase of emission current (H < -90 Oe) is presumably caused by the increase of applied magnetic field gradient and by rotation of the net magnetization of the probe off the probe axis towards the applied field direction. Lastly, the emission current does not normally return to $I_0$ when H is swept from $H_0$ to 0 Oe, but will recover to $I_0$ after a complete cycle of magnetic field sweeping. This may be understood as being caused by a certain degree of inelasticity of the probe under a large magnetic field, though more in-depth analyses are needed in order to reveal on the true behavior of this nanoelectromechanical system. What is of importance here is that these observations demonstrate strongly the excellent stability of field emission current from CNW emitters.

Fig. 6(d) shows the results of optimum fitting to the I-t curves corresponding to larger $H_0$ ($\geq$ 99.1 Oe) using Eq. (1). The two peaks are fitted separately in similar procedure described previously and with the same set of parameters except for α, which is weakly dependent on $H_0$ and in the range from $1.9 \times 10^4 - 3.1 \times 10^4$ nm Oe$^{-1}$. Comparisons between the extracted Δ$d$ and the simulation result using Eq. (2) show good agreements [inset of Fig. 6(d)]. In addition, it is found that the probe deflection in the maximum investigated magnetic field (148.1 Oe) is only ~ 2.5 nm. Even with such a small deflection, a large current modulation ratio ($I_{max}/I_{min}$) of over two orders of magnitude can be achieved [inset of Fig. 7(a)] at $d_0 = 11$ nm.

## *C. Scalability of dynamic control of field emission current*

To further explore the scalability of dynamic control of field emission current from CNW, similar field emission measurements were performed at different $d$ with constant $H_0 = 40.46$ Oe, which gives a probe deflection of ~ 0.7 nm. The normalized I-t relation in three continuous cycles of sinusoidal magnetic field sweeping is shown in Fig. 7(b). The superimposing lower and upper curves are I-t curves at a distance of 1.38 nm and 11 nm, respectively. It can be seen that the response of the emission current from 2D carbon to modulation is well reproduced in all three cycles, and shows a strong dependence on $d$. For the sake of clarity, Fig. 7(a) shows the dependence of the $I_{max}/I_{min}$ ratios on different $d$ (symbols) and the averaged $I_{max}/I_{min}$ ratio is shown as the solid curve as a visual guide. Clearly, the emission current modulation becomes more efficient at a smaller $d$, suggesting that dynamic control of local field emission current by varying $d$ is scalable in nanoscale field emission device applications.

## *D. Dynamic control of local field emission current with a superimposing AC voltage bias*

We next turn to modulating local field emission current with an AC electric voltage of variable amplitude superimposed on a DC bias. The amplitude of the AC electric field ($\Delta E_0 = 0.2 - 3.0$ kV/µm) is relatively small as compared to the typical bias field (27.6 kV/µm) at the investigated distance of ~1.3 nm. A current modulation ratio of $1.3 - 123$ has been achieved [Fig. 8(a)]. To have a more in-depth understanding of the modulation mechanism, Fig. 8(b) shows the typical experimental $I - t$ curves (symbols) corresponding to three different $\Delta E_0 = 0.5$, 1.6 and 2.6 kV/µm [dotted lines in Fig. 8(a)] together with their optimum fitting curves (dotted curves) using Eq. (1) with experimental parameters V = 38.1 V, $d_0$ = 1.32 nm, H = 0 Oe, β = 0.2 and S = ~500 nm$^2$. The upper two curves have been vertically shifted for clarity. At this point of discussion, it should be noted that β has been extracted from the slope of the F – N curve. The small value of β has been discussed in detail in Ref [32]. Apparently, variation of bias voltage alone is unable to fully account for the large current modulation observed at such a small $d$ of 1.32 nm, where effects arise from electrostatic interactions can be significant under a large electric field. The most likely explanation for the enhanced current modulation is that $d$ is modulated too due to capacitive effects between the probe and CNW [inset of Fig. 8(c)]. This argument is supported by the significant improvement in the agreement between the experimental data and the fitting curves (solid curves in [Fig. 8(b)] with Eq. (1) with H being replaced by the amplitude of the AC electric field (ΔE). To quantify the electrostatically induced probe deflection and further examine the understanding, the dependence of $\Delta d_E$ (defined as $d_0 \alpha \Delta E_0$) on $\Delta E_0$ is shown in Fig. 8(c) (symbols). Assuming a simple capacitor-and-cantilever model, the initial deflection of the probe under a macroscopic electric field $E_0$ and zero AC electric field (i.e. ΔE = 0) can be estimated as:

$$\Delta d_{E0} = \frac{A\varepsilon_0 E_0^2}{6E_m I^*}(\sin\theta\, L)^3, \tag{3}$$

where $\varepsilon_0$ is the absolute permittivity and $A = 3.48 \times 10^5$ nm$^2$ is the area of the bottom surface of the anode estimated from the probe size. During the derivation, both the anode surface and the CNW emitter have been assumed to be flat metallic surfaces extending to infinity to simplify the discussion. Taking $\Delta d_{E0}$ as a reference deflection, the net deflection caused by the superimposing AC electric field is given by:

$$\Delta d_E = \frac{A\varepsilon_0 E^2}{6E_m I^*}(\sin\theta\, L)^3 - \Delta d_{E0}, \tag{4}$$

where E is the total macroscopic electric field. The simulated dependence of $\Delta d$ on $\Delta E_0$ is shown as the solid curve in Fig. 8(c). It can be seen that the simulation result agree with the experimental data in the general trend for $\Delta E_0 = 0 - 2.3$ kV/μm, though the experimental $\Delta d$ increases much faster with the increase of $\Delta E_0$ for larger superimposing AC electric field ($\Delta E_0 > 2.3$ kV/μm). The reason for the latter observation is still not clear yet. The difference between experimental and simulated values is attributed to the large morphological and electrical difference between CNW and a flat metallic plate. Further systematic investigations are required in order to understand the true behavior of the electromechanical system involving the probe and the CNW in this regime.

## *E. Dynamic control of local field emission current from Fe/CNW with a W anode in an AC magnetic field*

In view of potential difficulty with the use of magnetic anode in certain applications, we have also investigated the possibility of emission current tuning using a magnetic field without resorting to a magnetic probe. To this end, the same the field emission measurements with an AC magnetic field described Section III (B) were repeated on Fe-coated CNW at d = 11 nm. A W anode was used instead of a Ni one so that it does not respond to the variation of the applied magnetic field.

Although thin metal coating will inevitably change the intrinsic field emission properties of CNW, it provides an alternative route for emission current modulation which may appear to be more attractive for certain applications. As shown in Fig. 9(a), the emission current decreases with increasing the magnetic field, indicating that *d* is increased by the elastic deformation of the Fe-coated CNW. Based on our previous work,[31,37] it is understood that the coated Fe has a large thickness near the edge. Therefore, when a magnetic field with gradient in the vertical direction is applied, the CNW will be deformed due to attractive (repulsive) interactions between the Fe layer at the top edge and the applied field. This will lead to a decrease (increase) of *d* and hence an increase (decrease) in the emission current [Fig. 1(c) and (f)]. Current modulation ratio up to only 4.3 was obtained for the maximum magnetic field amplitude investigated (i.e. 591 Oe) for a 5 nm thick Fe layer and with a 133 nm probe [Fig. 9(b)]. The relatively small modulation ratio is presumably caused by both the rigidness of CNW and wide spread of Fe on the CNW surface. This is further reflected in the fact that a thicker Fe layer (26 nm) reduces the modulation ratio, which is more obvious when the measurements were repeated with a larger probe (1.8 μm). This is because in addition to generating a magnetostatic force through interaction with the magnetic field, the Fe coating also increases the rigidity of CNW. It is worth noting that this is just a proof-of-concept experiment; a much larger modulation ratio is expected once the emitter structure is optimized including the ferromagnetic coating layer.

## IV. CONCLUSIONS

In summary, systematic experiments have been performed to modulate the local field emission current from CNW by varying the anode-cathode distance, and by varying the electric field in an UHV environment. Stable field emission current was obtained and current modulation ratio up to 105 and 123 has been achieved for the former and latter case, respectively. The experimental results have been explained by the F – N model in combination with a simple cantilever model to account for the change in either electric field or anode-cathode distance. Our results have demonstrated good stability of the local field emission current from CNW during the emission current modulation process and good scalability of current modulation at nanoscale, suggesting that CNW is a reliable emitter material for nanoscale field emission electronic devices. Although we have used a probe as the anode in this work, in practical applications, the CNW-probe configuration may also be replaced by a microelectromechanical system involving 2D carbons.

# ACKNOWLEDGEMENTS

This work is supported by the National Research Foundation of Singapore under Grants No. NRF-G-CRP 2007-05 and R-143-000-360-281, and Agency for Science, Technology and Research (A*STAR), Singapore, under Grant No. R-398-000-020-305.

**Figure Captions**

Figure 1. (Color online) A schematic diagram of dynamic control of field emission current from (a) bare CNW with a Ni anode in an AC magnetic field, (b) bare CNW with an AC electric field, and (c) Fe/CNW with a W anode in an AC magnetic field. The corresponding energy diagrams for (a), (b) and (c) are shown in (e), (f) and (g), respectively.

Figure 2. (Color online) A photo of the sample stage and schematic of the probe and electromagnet setup used in this work.

Figure 3. Schematic diagrams showing the process of calibrating the downward step size of probe on patterned gold features.

Figure 4. (Color online) (a) A 3-step schematic of the local electrical melting process. (b) Typical SEM images of probes as prepared and after the electrical melting process. All scale bars are 1 μm. (c) SEM image for local field emission measurements on CNW/Cu using a Ni probe as an anode at $d = 11$ nm. The lower inset is a close-up view of the as-grown CNW (scale bar: 500 nm).

Figure 5. (Color online) Typical field emission stability measurement with constant bias voltage at (a) large and (b) small emission current. The current compliant was set to 400 nA. Inset of (b) shows the W probe used for the measurement. (c) Dependence of the electric field required for 1 nA emission current on anode-cathode distance, obtained from CNW with probe of different sizes indicated in legend. (d) Typical response of the emission current to 15 cycles of AC magnetic field of different amplitudes ($H_0$).

Figure 6. (Color online) (a) Response of field emission current to one cycle of sinusoidal magnetic field of different $H_0$ at $d = 11$ nm. Color scale is normalized with respect to the emission current magnitude in zero magnetic field (t = 0 s). Dotted lines indicate the time when the emission current returns to its zero-H-field value. Superimposed with the color contour plot is the typical response of the emission current to a small (large) AC magnetic field in white (black). (b) and (e) are typical normalized I-H curves at small and large $H_0$, respectively. Black arrows indicate the sweeping direction of the magnetic field. Insets illustrate a simple cantilever model. (c) and (d) show the response of emission current (symbols) to small and large AC magnetic fields in I-t plot, respectively. Solid curves are the optimum fitting curves, and $H_0$ are indicated in unit of Oe beside the respective curves. Inset compares the maximum experimental probe deflection (symbols) with simulation results (solid line) at different AC magnetic fields. (f) Kinks in the I-H curves constantly observed before reversal of net magnetization of the Ni anode. $H_0$ is shown as figures beside the curves in unit of Oe.

Figure 7. (Color online) (a) Dependence of current modulation ratio on $d$ with $H_0 = 40.46$ Oe. Inset shows the current modulation ratio obtained in different $H_0$ at $d = 11$ nm. (b) Response of emission current to 3 continuous cycles of AC magnetic field ($H_0 = 40.46$ Oe) at different $d$. Color scale is normalized with respect to the emission current magnitude at t = 0 s. Typical

response of emission current to the magnetic field at a small (large) $d$ is shown as the superimposing lower (upper) curve.

Figure 8. (Color online) (a) Response of the field emission current to one cycle of sinusoidal electric field of different magnitude ($\Delta E_0$) superimposed on a constant DC bias field at $d = \sim 1.3$ nm. Color scale is normalized with respect to the emission current magnitude at t = 0 s. I-t curves with three typical $\Delta E_0$ (indicated by dotted lines) are shown in (b). Solid (dotted) solid curve is the fitting curve with (without) electrostatic interactions between the anode and CNW taken into considerations. (c) Experimental (symbols) and simulated (solid line) maximum electrostatically induced probe deflection at different $\Delta E_0$. Inset is a schematic of the capacitor-and-cantilever model.

Figure 9. (Color online) (a) Response of the field emission current from Fe (5 nm) coated CNW to one cycle of sinusoidal magnetic field of different $H_0$ at $d = 11$ nm. Color scale is normalized with respect to the emission current magnitude at t = 0 s. Typical response of the emission current to a small (large) AC magnetic field is shown as the superimposing dotted (solid) curve. (b) Current modulation ratio obtained from CNW coated with two different Fe layer thicknesses (5 and 26 nm) and using W probes of two different sizes (0.13 and 1.8 µm) as an anode.

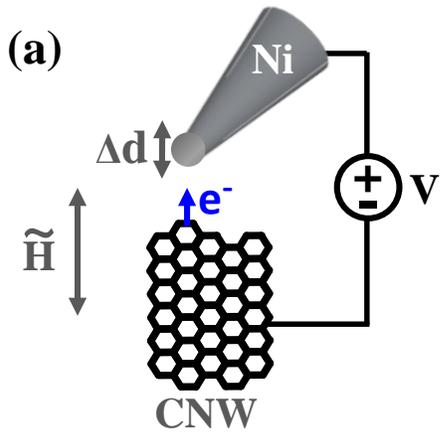
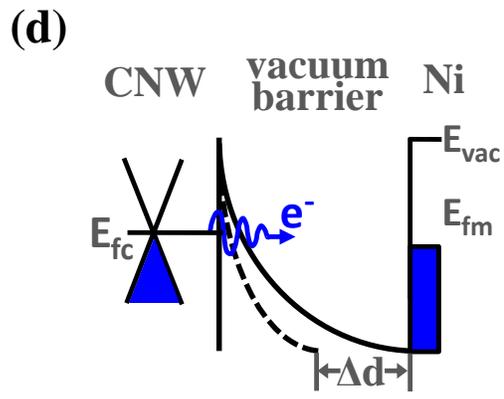
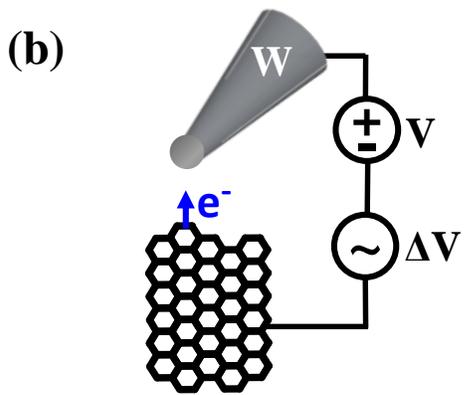
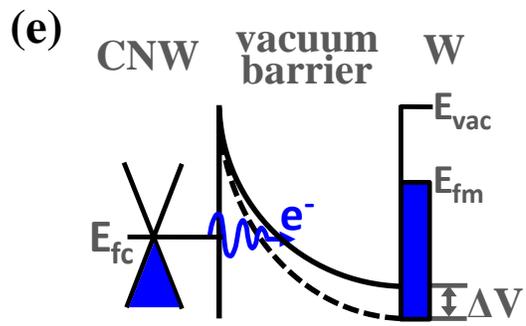
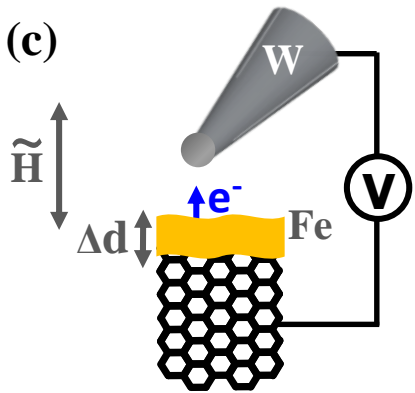
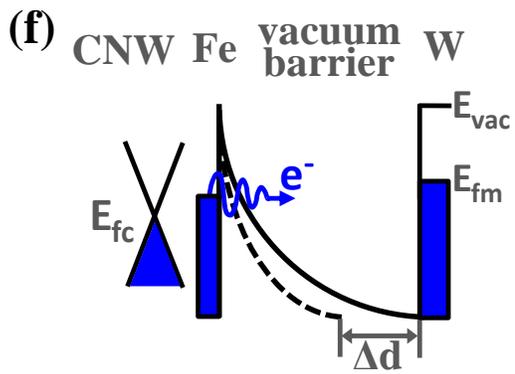

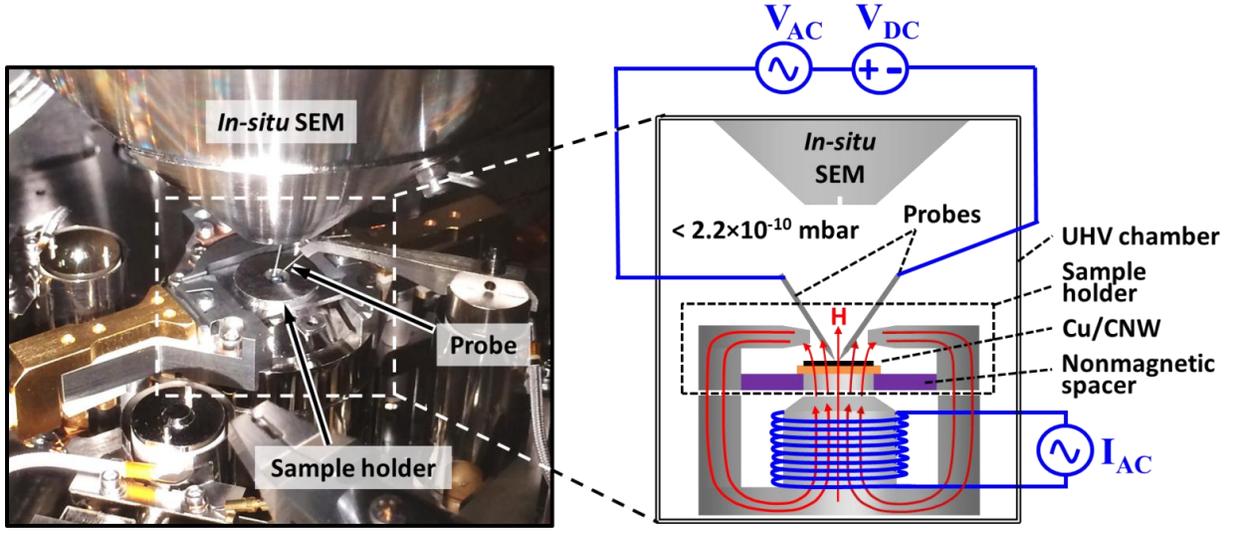

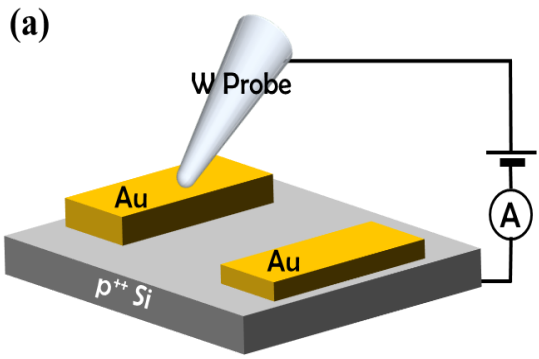 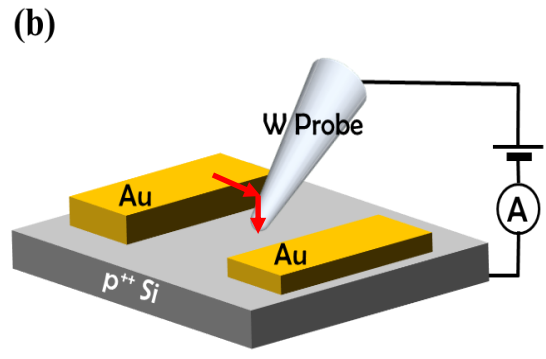

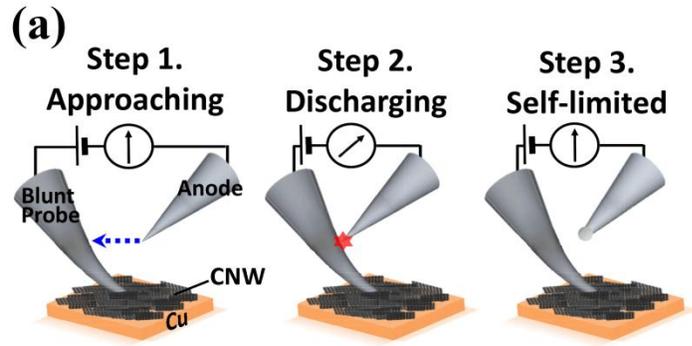

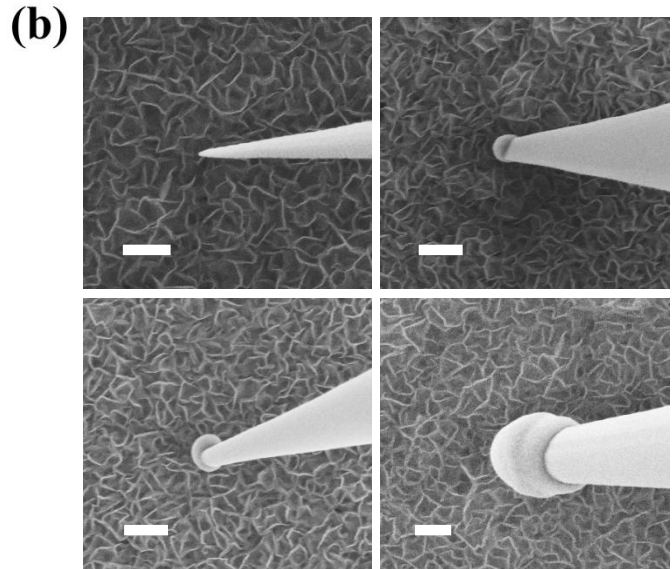

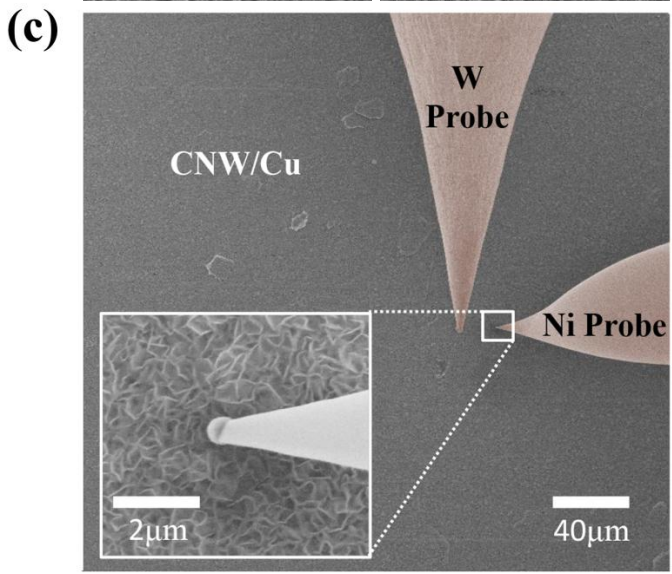

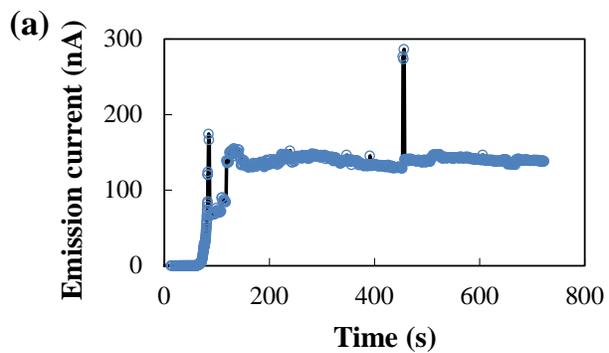
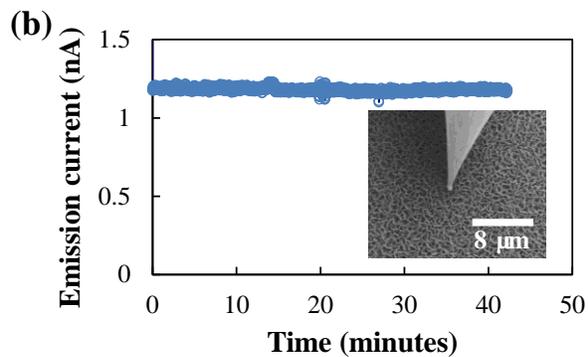
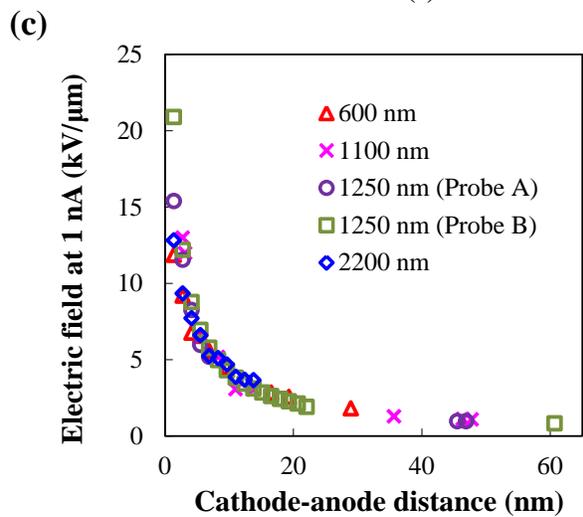
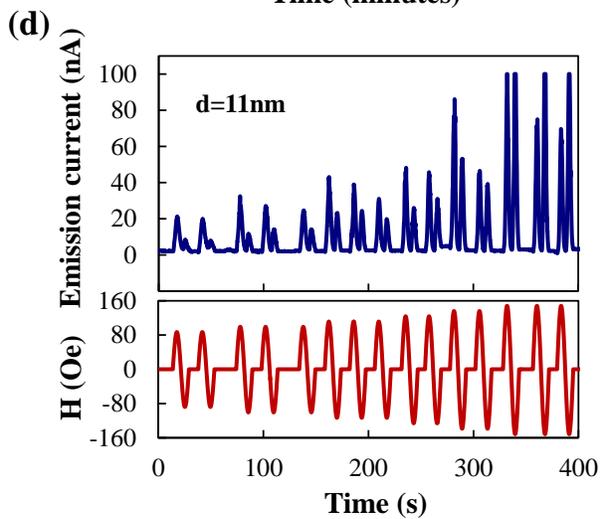

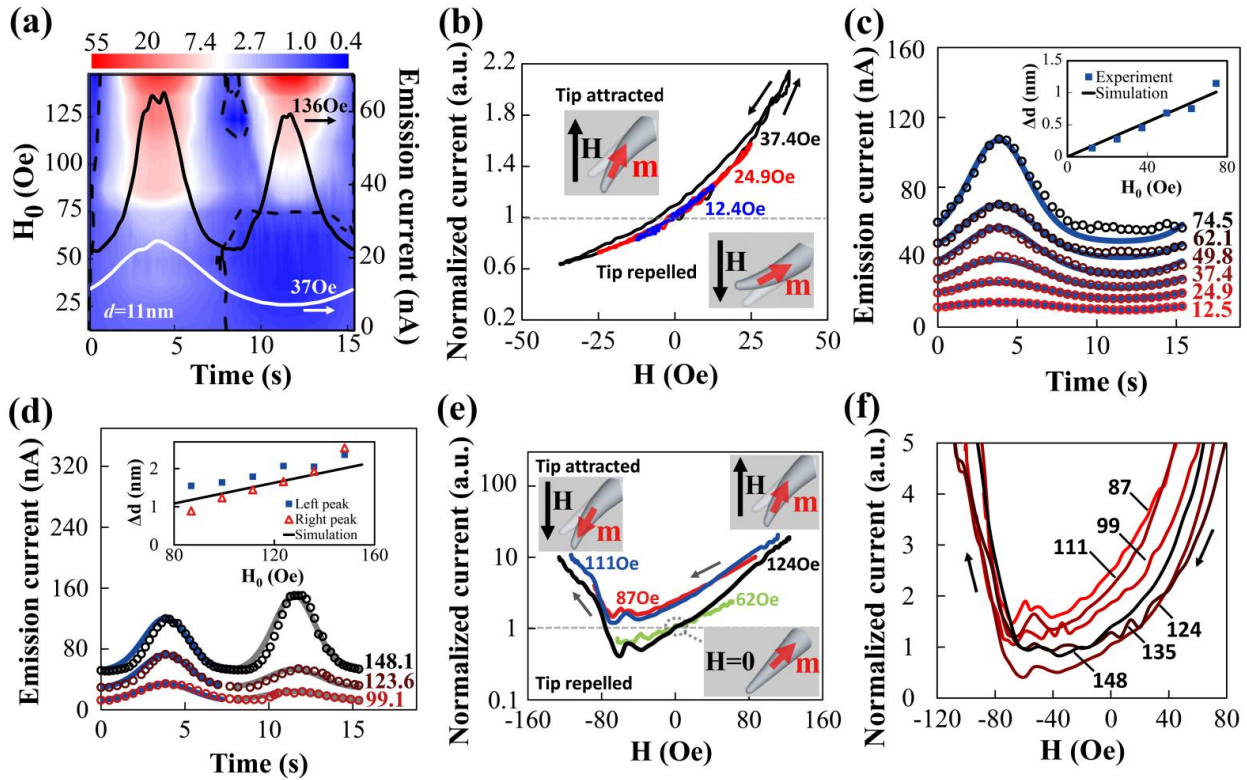

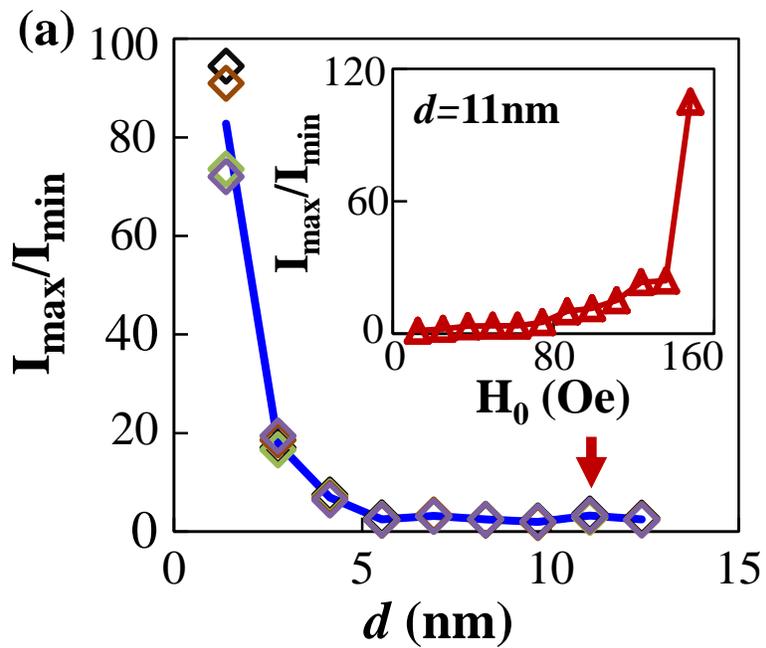
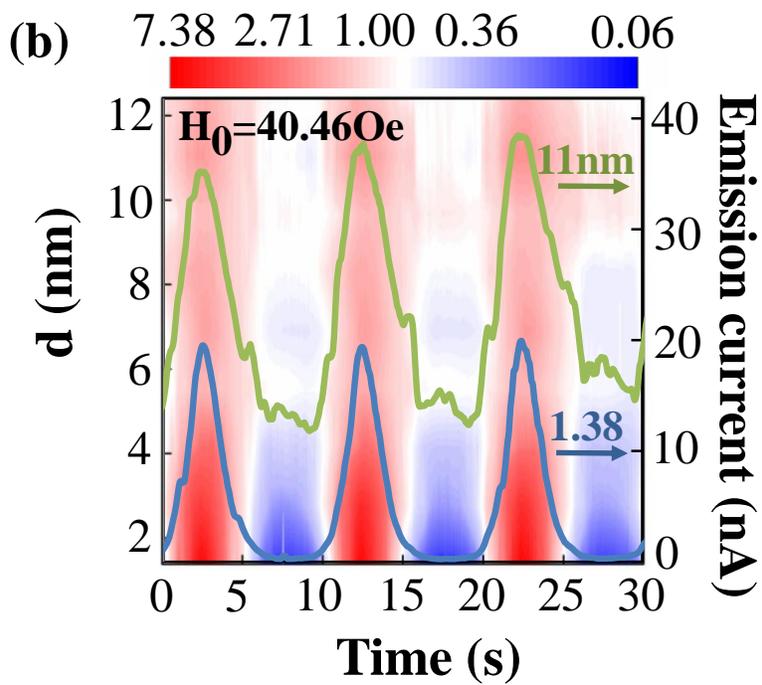

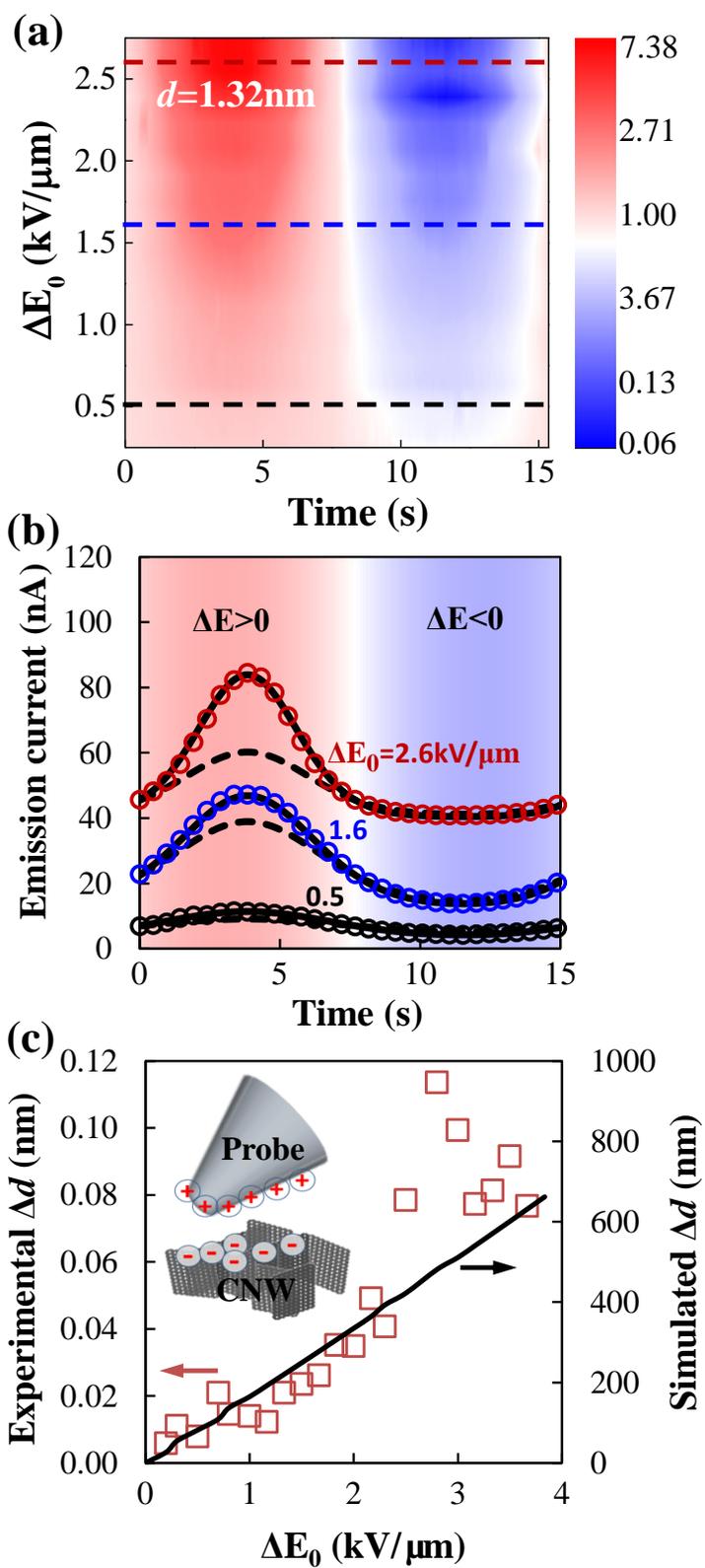

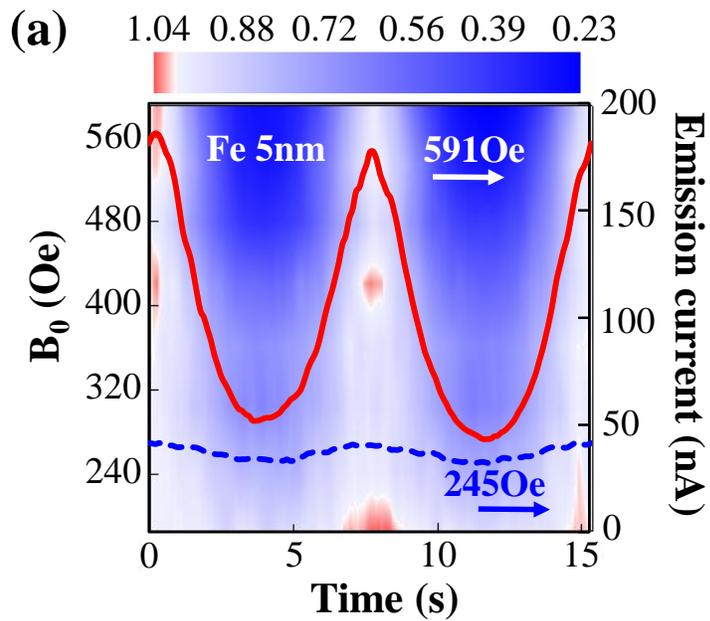

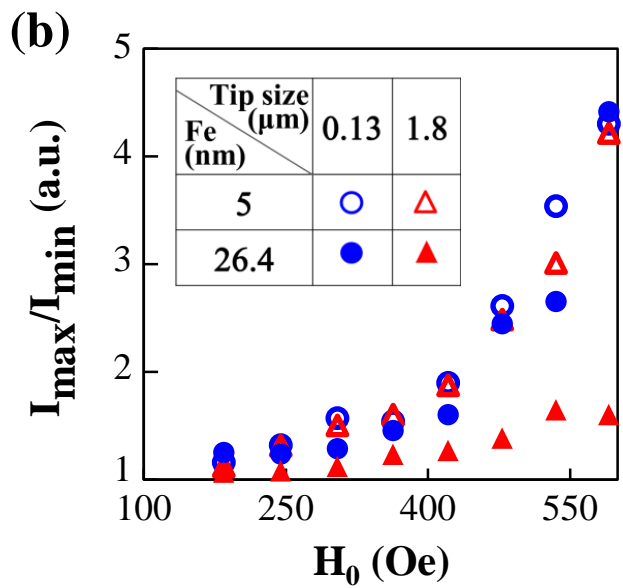